\documentclass[namedreferences,hyperref,optionalrh,solaromanenum]{spr-sola}

\usepackage{graphicx}                    
\usepackage{amssymb}                    
\usepackage{color}                       

\newcommand{\apss}{{\it Astrophys. Space Sci.}}

\newcommand{\solphys}{{\it Sol. Phys.}}

\chardef\us=`\_

\begin{document}

\begin{frontmatter}

\title{Measurement of Solar Differential Rotation 
by Absolutely Calibrated Iodine-Cell Spectroscopy}

%
\author[email={ytakeda@js2.so-net.ne.jp}]{\inits{Y.}
\fnm{Yoichi}~\snm{Takeda}\orcid{0000-0002-7363-0447}}
\address{11-2 Enomachi, Naka-ku, Hiroshima-shi 730-0851, Japan}


\runningauthor{Y. Takeda}
\runningtitle{Solar Differential Rotation and Iodine-Cell Spectroscopy}

\begin{abstract}
The iodine-cell technique, which is known to be efficient in 
precisely establishing Doppler velocity shifts, was once applied 
by the author to measuring the solar differential rotation based 
on full-disk spectroscopic observations (Takeda and Ueno, 
{\it Sol. Phys.} {\bf 270}, 447, 2011). 
However, the data reduction procedure (in simple analogy with 
the stellar case) adopted therein was not necessarily adequate, 
because specific characteristic involved with the disk-resolved Sun 
(i.e., center--limb variation of line strengths) was not properly 
taken into consideration. 
Therefore, this problem is revisited based on the same data but 
with an application to theoretical spectrum fitting, which can 
yield absolute heliocentric radial velocities ($v_{\rm obs}$) 
in a consistent manner as shown in the study of solar gravitational 
redshift (Takeda and Ueno, {\it Sol. Phys.} {\bf 281}, 551, 2012). 
Likewise, instead of converting 
$v_{\rm obs}$ into $\omega$ (angular velocity) at each disk point, 
which suffers considerable errors especially near the central meridian, 
$\omega$ was derived this time by applying the least squares analysis 
to a dataset comprising $v_{\rm obs}$ values at many points.
This new analysis resulted in $\omega$ (deg~day$^{-1}$) = $13.92 (\pm 0.03)$ 
$-1.69(\pm 0.34)\sin^{2}\psi$ $-2.37(\pm 0.62) \sin^{4}\psi$ ($\psi$: 
the heliographic latitude) along with the gravitational redshift of 675~m~s$^{-1}$, 
which are favorably compared with previous publications. In addition,
how the distribution of observing points on the disk affects 
the result is also examined, which reveals that rotation parameters
may suffer appreciable errors depending on cases.
\end{abstract}

%

\end{frontmatter}

%

\section{Introduction}

\subsection{Past Studies of Solar Differential Rotation}

Ever since Christoph Scheiner noticed about 400 years ago from sunspots 
observations that the Sun rotates with a shorter period in the equator 
than in higher latitudes, a number of observational studies on this 
``solar differential rotation'' were conducted especially in the 20th century 
(mainly by invoking two approaches of sunspot tracing and spectroscopic 
Doppler method), and various information on its characteristics has 
been accumulated. Moreover, the advent of solar seismology has enabled
to diagnose the nature of internal rotation successfully.
See, e.g., Patern\`{o} (2010) for a review concisely summarizing 
the historical aspect and current status on this subject. 

Nowadays, observationally studying the surface differential rotation 
on the visible solar disk is rather classical and not the mainstream 
of solar physics, given that the main features had been established 
in 1970s\,--\,1980s as reviewed by Schr\"{o}ter (1985). 
In the author's opinion, however, it is still important and worth 
pursuing to improve the precision of the rotational parameters as much 
as possible, in order to clarify their dependence upon the solar 
activity cycle since both should be closely connected through the
dynamo mechanism.

From the viewpoint of accomplishing higher accuracy, the spectroscopic 
approach (quickly yielding the result and potentially applicable 
to any higher latitude) may be superior to the tracing method 
(which requires a rather long monitoring and is limited up to a 
certain latitude depending on the activity phase).  

Since radial velocities\footnote{Following the usual astronomical convention,
we use the term ``radial velocity'' for the velocity component along the 
line of sight (positive for the direction towards recession) in this article. 
Note, however, that this terminology is not so widely used in solar physics
(for which ``line-of-sight velocity'' may be more preferred), presumably 
because the adjective ``radial'' is rather confusing in distinguishing from
the meaning of center-to-limb direction on the solar disk.
} at many points on the solar disk have to be efficiently determined 
in this case (for which classical manual work on the spectra  
is hardly practicable), the best choice would be the Doppler compensator method 
by using a magnetograph (e.g., Howard and Harvey, 1970). Actually,
most of the spectroscopic measurements of solar differential rotation
over the past half century (mainly in 1960s\,--\,1980s) have employed this 
technique (see, e.g., Schr\"{o}ter, 1985, and the references therein).
 
But unfortunately, this method is not necessarily easy to practice 
from the technical as well as budgetary point of view, because it 
requires an instrument manufactured with skill and carefully tuned. 
Accordingly, application of this technique to solar differential 
rotation measurement was experienced mainly in large observatories 
(e.g., Mt. Wilson or Kitt Peak).  

\subsection{Iodine Cell as an Effective Spectroscopic Tool} 

Here, the iodine-cell spectroscopy may serve as a promising 
alternative, because it also enables fairly precise measurement of 
Doppler shifts based on efficient data reduction procedures using 
an optimization algorithm. This method became very popular in application 
to searching for extrasolar planets around stars, although it was originally 
introduced in solar physics. The decisive merit of this technique 
is its simplicity and cost effectiveness: what should be done is 
only to place the gas filter containing iodine vapor somewhere 
in the light path of the spectrograph. 

A trial of applying the iodine-cell technique to measurement 
of solar differential rotation was done by Takeda and Ueno (2011; 
hereinafter referred to as Paper~I) based on full-disk spectroscopic
observations, which yielded results more or less reasonable.
However, some unsatisfactory drawbacks are noticed in the data 
reduction procedures adopted in that work as described below.

\subsubsection{Disregarded Center--Limb Spectral Variation}

Firstly, an alarming weakpoint was the choice of the reference spectrum. 
The essence of data analysis in the iodine-cell method is to simulate 
the ``object+iodine'' spectrum (to be compared with the observed one)
based on the (i) ``pure iodine'' spectrum and (ii)  ``pure object'' 
(reference) spectrum appropriately Doppler-shifted. In Paper~I, 
the solar disk-center spectrum was adopted as the representative 
reference spectrum throughout the analysis (see Section~3 therein). 
This assumption means that the center--limb spectral variation, 
differing from line to line as extensively studied by Takeda and UeNo 
(2019), was neglected, which must have lead to some imperfect match 
between the simulated and observed spectra (especially near to the limb). 
This is presumably the reason for the fact that the resulting radial 
velocities fluctuated and their errors were appreciably position-dependent.

A remedy for this problem (in order to appropriately handle disk-resolved 
solar spectra by correctly taking into account the center--limb variation)
is to use ``theoretical reference spectra'' which are so modeled as to  
well reproduce observed solar spectra by adequately adjusting the 
parameters (abundances, line broadening, etc.). This approach was
soon later adopted by Takeda and Ueno (2012; hereinafter Paper~II) in the
trial of detecting the solar gravitational redshift, since ``absolute''
radial velocities are obtained in this case. Although the precision itself
attained in such an absolute analysis is quantitatively lower as compared 
to the case of purely relative analysis (as done in Paper~I), it has 
a merit that errors in the velocity solutions are guaranteed to be
consistently uniform, which should be more essential in the present context. 
Accordingly, it is worth reinvestigating the nature of solar differential 
rotation based on the model-based analysis as done in Paper~II.  
Since heliocentric radial velocities in the absolute scale are usable
in this case, there is no need to make use of the symmetry about the 
central meridian (which has been adopted in almost all solar rotation studies)
and the gravitational redshift can be simultaneously derived 
as a by-product based on the full-disk data.

\subsubsection{Inadequate Derivation of Angular Velocity}

Secondly, regarding the evaluation of the rotational angular velocity ($\omega$) 
as a function of heliographic latitude ($\psi$) based on the observed 
radial velocities ($v_{\rm obs}$) at each of the many points $(x,y)$ 
on the disk, the procedure adopted in Paper~I was not necessarily adequate. 
That is, $\omega_{i}(\psi_{i})$ was ``directly'' derived from $v_{{\rm obs},i}$ 
at each individual point $(x_{i},y_{i})$, because $\omega$ is analytically
related to $v_{\rm obs}$. The final $\omega$ versus $\psi$ relation 
(parameterized as a second-order polynomial in terms of $\sin^{2}\psi$)
was determined from the least squares analysis based on the resulting 
set of $\omega_{i}(\psi_{i})$ $(i=1, 2, \ldots, N)$.

The problematic point is, since $\omega (\propto v_{\rm obs}/x)$ is almost inversely 
proportional to $x$ (distance from the meridian along the east--west direction), 
even small errors involved in $v_{\rm obs}$ are considerably enhanced 
when converted to $\omega$ if $|x|$ is small. As a matter of fact, those 
$\omega_{i}$ data at $|x_{i}| < 0.3 R$ ($R$ is the solar radius) had to be 
discarded in Paper~I, 
which was a big waste because more than half of the total points could 
not be used.  Besides, $\omega_{i}$ data of diversified error sizes
had to be combined, which was not sound.

A much better way of analysis should be to determine $\omega(\psi)$ from 
the ``ensemble'' of $v_{\rm obs}$ data themselves (instead of directly converting 
$v_{\rm obs}$ to $\omega$ at each point) by applying the least square analysis, 
such as previously done by Howard and Harvey (1970). Therefore, we analyze 
the radial velocity data by following this approach, where two different
cases are examined: (a) $\omega(\psi)$ is expressed by a quadratic polynomial 
in terms of $\sin^{2}\psi$ and their parameters are determined based on all 
$v_{\rm obs}$ data, and (b) $\omega$ is determined from the $v_{\rm obs}$ data 
belonging to each $\psi$ bin of $10^{\circ}$ wide (without postulating any 
analytic expression of $\omega$ versus $\psi$ relation). It would be interesting
to check where both results satisfactorily agree with each other.  

\subsection{Objectives of This Study}

Accordingly, being motivated to overcome the problems involved in Paper~I, 
the purpose of this investigation is (i) to reanalyze 
the same observational data to yield absolute $v_{\rm obs}$ by the 
model-based procedure employed in Paper~II, and (ii) to derive $\omega$ 
as a function of $\psi$ by applying the least squares analysis to the 
data set of $v_{\rm obs}$. 

Also, as a by-product of the present analysis of full-disk radial velocities, 
the solar gravitational redshift should be obtained. It is interesting to 
compare this value with that obtained in Paper~II, which was obtained only 
from the meridian data free from rotation. 

In addition, as a related application, we study how the distribution  
of observed points on the disk (adopted for the analysis) affects the results
by examining various test cases, which would be meaningful for understanding 
the critical factor(s) influencing the precision of rotational parameters. 
This is another aim of this article. 
  
\section{Observational Data}

The observational data used in this study are the same as adopted
in Paper~I and Paper~II. These high-dispersion spectra covering 5188\,--\,5212~\AA\ 
(imprinted with lines of I$_{2}$ molecules) were obtained by two sets of 
full-disk covering observations (on 20\,--\,21 July 2010, when the rotational
axis is tilted by $B_{0} = +4.8^{\circ}$) done with east--west aligned 
slits (A-set) and north--south aligned slit (B-set) by using the Domeless Solar 
Telescope at Hida Observatory of Kyoto University. Each long-slit spectrum 
was spatially divided into three sub-spectra designated by [l, c, r] 
(for the A-set) or [t, m, b] (for the B-set). Therefore, two characters 
(such as ``Al'' or ``Bm'') are assigned to each (sub-)spectrum to discern 
the corresponding spacial position (relative to the slit center) on the disk.  
Since the scanning over the disk was done with $\Delta r = R/12$ (step of 
radial direction) and $\Delta \theta = 360^{\circ}/48$ (step of position angle) 
for both sets, spectra at 3456 ($= 2 \times 12 \times 48 \times 3$) points\footnote{
This is the gross number of the observed points in total. There are cases 
that they happen to positionally overlap with each other.} 
covering $r(=\sqrt{x^{2}+y^{2}}) =0$ 
(disk center) to $r = 0.961 R$ (near to the limb, $R$ is the solar radius) 
are eventually available for the analysis.\footnote{The spectra observed at 
the extreme limb (designated by suffix `12' such as $r_{12}$ in Paper~I) are 
not used in this investigation (as in Paper~II) because of their lower reliability.}  
The corresponding locations of all these spectra on the disk are illustrated
in Figure~1. See also Section~2 of Paper~I for more detailed descriptions 
regarding these data. 

\setcounter{figure}{0}
\begin{figure}
\centerline{\includegraphics[width=6.0cm]{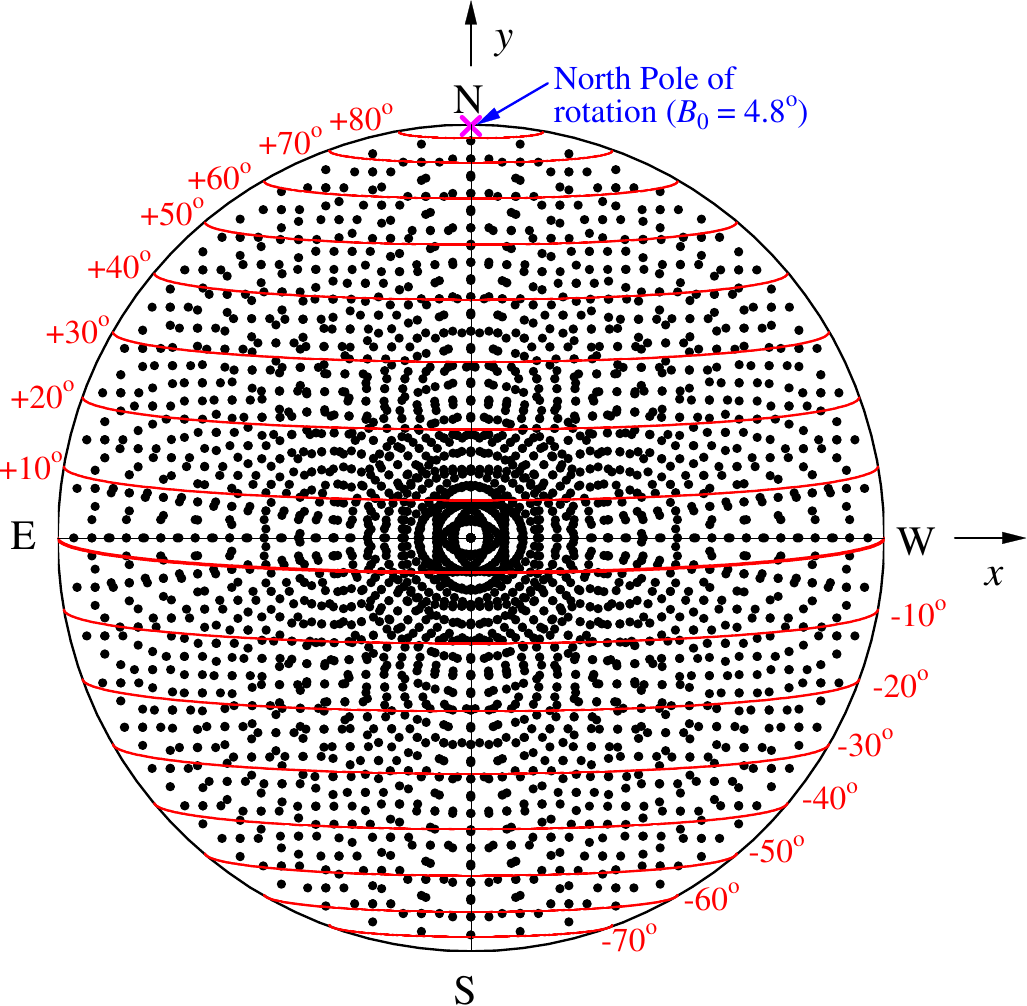}}
\caption{
Distribution of observed points on the solar disk, at which  
radial velocities are determined. The loci of constant 
$\psi$ (heliographic latitude) are also depicted by red lines.
}
\label{fig1}
\end{figure}

\section{Radial Velocity Determination} 

The procedures of deriving the absolute heliocentric radial velocities ($v_{\rm obs}$) 
from the observed spectra at each point of the solar disk are essentially the
same as adopted in Paper~II (see Sections~3\,--\,5 therein).
The only difference is that the center--limb variation of the solar photospheric 
microturbulence ($\xi$) is taken into account (instead of the constant value of 
$\xi$ = 1~km~s$^{-1}$ assumed in Paper~II) in fitting the observed spectra 
with theoretically modeled ones. That is, according to the recent work of Takeda (2022), 
\begin{equation}
\xi  = 1 + 0.0649 (1- \mu) + 1.427 (1 - \mu)^{2} \;\; ({\rm km \; s}^{-1})
\end{equation} 
is adopted (see Equation~1 therein). Here, $\mu$ is the direction cosine 
($\mu = \cos\Theta$: $\Theta$ is the angle between the line of sight and 
the surface normal), expressed in terms of ($x$, $y$) coordinates as
\begin{equation}
\mu = \sqrt{1 - (x^{2}+y^{2})/R^{2}}.
\end{equation}

The raw radial velocity ($V_{\rm r}^{\rm raw}$) at each point (along with 
the corresponding mean error $\epsilon$) is determined as the mean of the 
Doppler shifts (after being corrected for offset errors) calculated for 
four segments of 6~\AA\ wide (cf. Equations~9\,--\,11 in Paper~II). 

Then, the heliocentric radial velocity ($V_{\rm r}^{\rm hel}$) is derived by 
applying the heliocentric correction  ($\Delta^{\rm hel}$; see Section~4.1 
in Paper~I) as $V_{\rm r}^{\rm raw} + \Delta^{\rm hel}$.

Finally, we subtract the correction for the convective blue shift 
($\langle \delta V \rangle$, cf. Equation~17 in Paper~II) from $V_{\rm r}^{\rm hel}$, 
which is expressed by the following relation 
\begin{equation}
\langle \delta V \rangle = -278.5 + 79.9 (1-\mu) + 422.3 (1-\mu)^2 \;\; ({\rm m \; s}^{-1}),
\end{equation}
in order to obtain the final absolute radial velocity ($v_{\rm obs}$). 
Consequently, $v_{\rm obs}$ is derived as  
\begin{equation} 
v_{\rm obs} = V_{\rm r}^{\rm raw} + \Delta^{\rm hel} - \langle \delta V \rangle.
\end{equation} 

The resulting values of $V_{\rm r}^{\rm raw}$, $\Delta^{\rm hel}$, 
$\langle \delta V \rangle$, and $v_{\rm obs}$ at each of the 3456 observed points
on the disk are presented in the supplementary online material (tableE.dat).

The histograms for the distribution of errors ($\epsilon$, given by Equation~11 in 
Paper~II) involved with $v_{\rm obs}$ at each radius bin are depicted in Figure~2a, 
which shows that no significant dependence exists upon the position on the disk.  
Actually, their mean value ($\langle \epsilon \rangle$) obtained by averaging 
$\epsilon$ is almost constant at $\approx$~80\,--\,90~m~s$^{-1}$ irrespective of $r$ (Figure~2b),
in contrast to the case of Paper~I (see Figure~9b therein) where a systematic
increase of $\langle \epsilon \rangle$ with $r$ toward the limb was observed. 
Note also that the $r$-independent nature of $\langle \epsilon \rangle$ shown 
in Figure~2b is favorably more distinct than that in Figure~7b in Paper~II 
(essentially the similar figure), which is presumably due to the reasonable 
consideration of center--limb variation for $\xi$ in the present analysis.

\setcounter{figure}{1}
\begin{figure}
\centerline{\includegraphics[width=10.0cm]{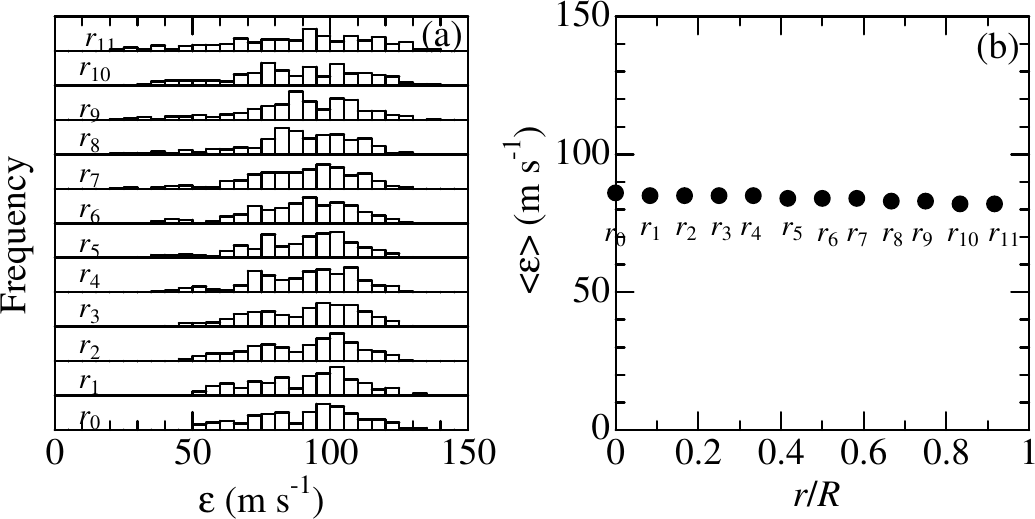}}
\caption{
(a) Distribution histogram for the errors ($\epsilon$) involved 
in observed radial velocities at each radius 
($r_{0}$, $r_{1}$, $\ldots$, $r_{10}$, $r_{11}$ 
corresponding to $r/R$ = 0, 1/12, $\ldots$ $10/12, 11/12$).
(b) Averaged errors ($\langle \epsilon \rangle$) calculated 
at each of the radius bins are plotted against the radius.
}
\label{fig2}
\end{figure}

Admittedly, the size of $\epsilon$ itself in the present case of absolute analysis
is several times larger in comparison to the relative analysis adopted in Paper~I,
because various factors\footnote{For example, since the wavelengths of 
spectral line data adopted in this analysis have precisions only to the third  
decimal in \AA, this already can be a source of uncertainty amounting to
several tens m~s$^{-1}$  (0.001~\AA\ corresponds to $\approx$~50\,--\,60~m~s$^{-1}$). }
 are involved as mentioned in Section~6 of Paper~II. However, as long as the 
present purpose (investigating the global feature of rotational velocity field)
is concerned, warranting homogeneous precision of $v_{\rm obs}$ over the disk 
should be more significant.

\section{Trends of Absolute Radial Velocities on the Disk}

It is worth examining the global characteristics of the resulting 
$v_{\rm obs}$ in terms of the symmetry with respect to the meridian.
In our choice of observed points on the disk, any point in the western 
hemisphere ($x^{+}(>0),y$) has its symmetric counterpart 
in the eastern hemisphere ($x^{-}(<0),y$), where $x^{+} + x^{-} =0$. 
Let us denote the radial velocities corresponding to the former and 
the latter points as $v_{\rm obs}^{+}$ and $v_{\rm obs}^{-}$, respectively.
The four panels of Figure~3 illustrate the correlations of 
(a) $v_{\rm obs}$ versus $x$, (b) $v_{\rm obs}^{+}$ versus 
$v_{\rm obs}^{-}$, (c) $0.5\,(v_{\rm obs}^{+} - v_{\rm obs}^{-})$ versus
$0.5\,(x^{+}-x^{-})$, and (d) $0.5\,(v_{\rm obs}^{+} + v_{\rm obs}^{-})$ versus $y$.
We can see the following trends by inspecting these figures. 
\begin{itemize}
\item
Generally, $v_{\rm obs}$ tends to increase with $x$, and $v_{\rm obs}^{+}$ 
and $v_{\rm obs}^{-}$ are inversely correlated with each other.
\item
The sum of $0.5\,(v_{\rm obs}^{+} + v_{\rm obs}^{-})$ is nearly constant, 
and their average over all pairs is 0.686~km~s$^{-1}$
(with a standard deviation of 0.081~km~s$^{-1}$).
\item 
This is interpreted as an offset to $v_{\rm obs}$ corresponding to the gravitational 
redshift. If this offset is subtracted, $v^{+} + v^{-} \approx 0$ is realized,
as expected for rotational velocities. 
\item
Accordingly, the global feature of $v_{\rm obs}$ is explained mostly by 
the rotational velocity and an offset constant, which serves as a reference 
in parameterizing the theoretical radial velocity in Section~5.1.
\end{itemize}

\setcounter{figure}{2}
\begin{figure}
\centerline{\includegraphics[width=8.0cm]{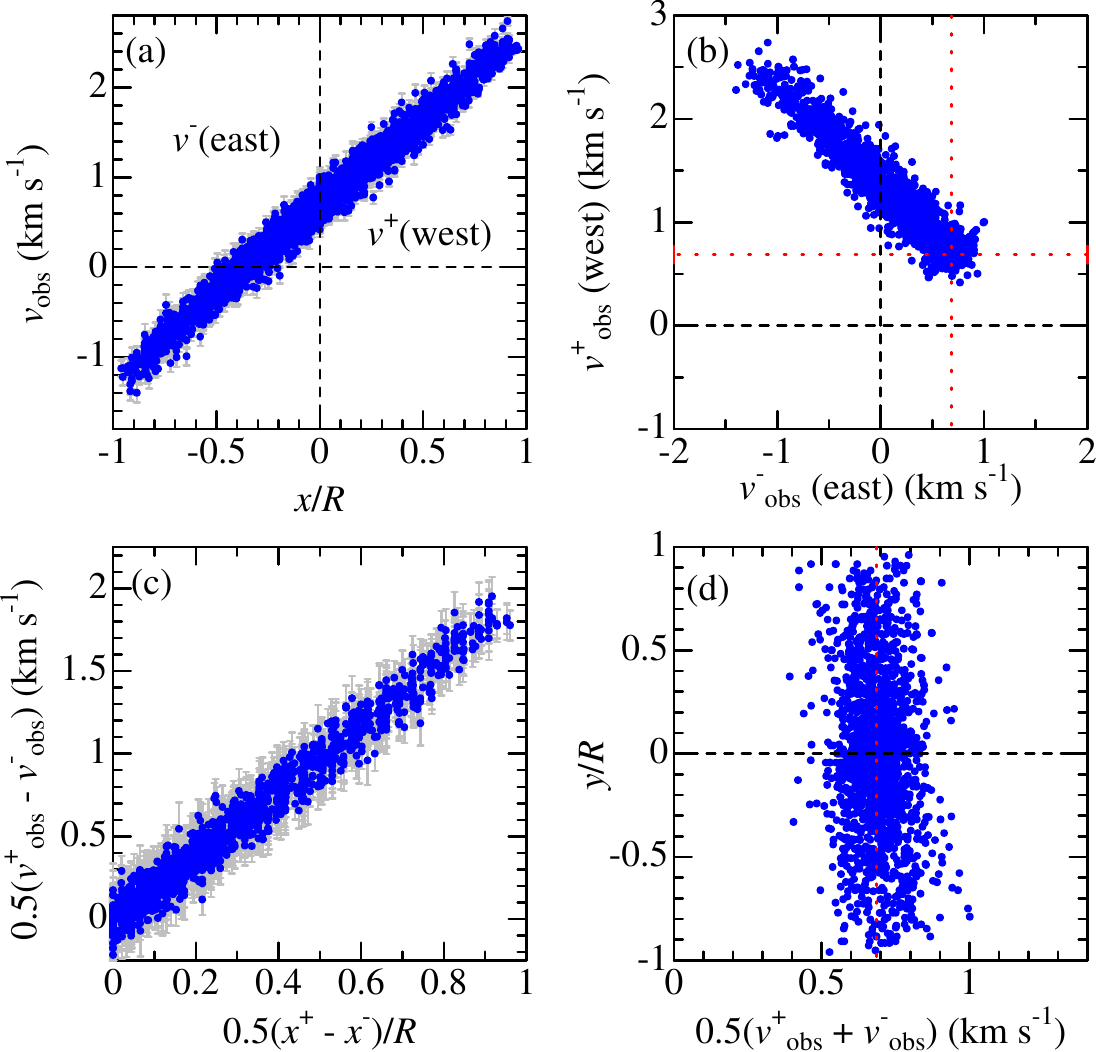}}
\caption{Trends of $v_{\rm obs}$ (observed absolute radial
velocities) in terms of the heliographic coordinates ($x$ and $y$).
Those in the eastern ($x<0$) and western ($x>0$) hemispheres 
are denoted by superscripts ``-'' and ``+'', respectively.
(a) $v_{\rm obs}$ versus $x$. (b) $v_{\rm obs}^{+}$ versus $v_{\rm obs}^{-}$.
(c) $0.5\,(v_{\rm obs}^{+} - v_{\rm obs}^{-})$ versus 
$0.5\,(x^{+} - x^{-})$, where ($v_{\rm obs}^{+}$, $v_{\rm obs}^{-}$)
and ($x^{+}$, $x^{-}$) are the reflectionally symmetric pairs
with respect to the meridian. 
(d) $0.5\,(v_{\rm obs}^{+} + v_{\rm obs}^{-})$ versus $y$.
In panels b and d, the position of 0.686~km~s$^{-1}$, which 
is the average of $\langle 0.5\,(v_{\rm obs}^{+} + v_{\rm obs}^{-})\rangle$
corresponding to the gravitational redshift, is indicated by the red
dotted line. In panels a and c are also shown the error bars in gray.
}
\label{fig3}
\end{figure}

\section{Derivation of the Angular Rotational Velocity}

\subsection{Case of Assuming an Analytical $\omega$ Versus $\psi$ Relation}

As mentioned in Section~4, it is reasonable to include only the rotational 
velocity component ($V_{\rm rot}$) plus an offset constant (corresponding to
the gravitational redshift) in modeling the absolute heliocentric radial 
velocity to be compared with $v_{\rm obs}$.
This implicitly assumes that any locally-fluctuating irregular velocity fields 
(e.g., supergranules or 5-minutes oscillation) are eventually averaged to become 
insignificant by using many points on the disk and the global velocity field 
(e.g., meridional circulation) is quantitatively negligible in comparison 
with rotation.  

The rotational velocity component along the line of sight $V_{\rm rot}$
at the heliographic longitude and latitude of ($\phi, \psi$)\footnote{ 
The range of $\phi$ is $-180^{\circ} \le \phi \le +180^{\circ}$, where 
$\phi = 0^{\circ}$ at the meridian (east/west limb is $-90^{\circ}/+90^{\circ}$). 
The range of $\psi$ is $-90^{\circ} \le \psi \le +90^{\circ}$, where 
$\psi = 0^{\circ}$ at the equator (south/north pole is 
$-90^{\circ}/+90^{\circ}$).} is written as
\begin{equation}
V_{\rm rot}(\phi,\psi) = R \omega(\psi) \cos\psi \sin\phi \cos B_{0}, 
\end{equation}  
where $\omega(\psi)$ is the angular velocity at latitude $\psi$.
The ($\phi, \psi$) spherical coordinate system is related to the ($x, y$) 
Cartesian coordinate system on the disk by the following equations.
\begin{eqnarray}
x & = & R \cos\psi \sin\phi \\ 
R \sin \psi & = & \sqrt{R^{2}-x^{2}-y^{2}} \sin B_{0} + y \cos B_{0}  
\end{eqnarray}
Therefore, from Equations~5 and 6,  $V_{\rm rot}$ is also expressed 
in the ($x, y$) system as
\begin{equation}
V_{\rm rot}(x, y) = \omega(\psi) x \cos B_{0},
\end{equation}
which indicates that $V_{\rm rot}$ is nearly proportional to $x$
if the weak dependence of $\omega$ upon the latitude $\psi$ is disregarded. 
 
As to the latitude dependence of $\omega$, we adopt a second-order 
polynomial in terms of $\sin^{2}\psi$ as usually done,
\begin{equation}
\omega(\psi) = A + B \sin^{2}\psi + C \sin^{4}\psi,
\end{equation}  
where $A$, $B$, and $C$ are constants. 

Therefore, by inserting Equation~9 into Equation~5, the $\chi^{2}$
to be minimized by the least squares analysis is expressed as
\begin{equation}
\chi^{2} = \sum_{i=1}^{N}\Biggl[\frac{v_{{\rm obs},i} - x_{i}\cos B_{0}(A + B \sin^{2}\psi_{i} + C \sin^{4}\psi_{i}) -D}{\sigma_{i}}\Biggr]^{2}, 
\end{equation}
where $v_{{\rm obs},i}$, $x_{i}$, $\psi_{i}$, and $\sigma_{i} (\equiv \epsilon_{i})$  
correspond to each point $i$ ($i = 1, 2, \ldots, N$), and $D$ is an offset constant.

The determination of four parameters ($A$, $B$, $C$, and $D$) minimizing 
$\chi^{2}$ given by Equation~10 was done by following the standard procedure of 
least squares analysis (e.g.,  Bevington and Robinson, 2003), which resulted in\footnote{
According to the convention, $A$, $B$, and $C$ are expressed in unit of
deg~day$^{-1}$, which are obtained by multiplying the values in unit of s$^{-1}$
(=~km~s$^{-1}$/km; since $v_{\rm obs}$ and $x$ are in units of km~s$^{-1}$ and km,
respectively) by $(180/\pi) \times 24 \times 60 \times 60$.}
$A = 13.92 (\pm 0.03)$,  $B = -1.69(\pm 0.34)$, and $C = -2.37(\pm 0.62)$ 
(deg~day$^{-1}$), while $D = 0.675 (\pm 0.001)$ (km~s$^{-1}$).
The corresponding $\omega$ versus $\psi$ relation is depicted in Figure~4.

These $A$, $B$, and $C$ values are almost consistent with those derived in
Paper~I [($A$, $B$, $C$) = ($14.03\pm 0.06$, $-1.84\pm 0.57$, $-1.92\pm 0.85$)]
within expected uncertainties, which means that the impact of improvements
in this investigation turned out not so remarkable after all. 
They are also well compared with most results reported in the past publications 
(see Figure~12 in Paper~I), such as that obtained in Mt. Wilson Observatory
(Howard and Harvey, 1970) also shown in Figure~4 for comparison. 

Regarding the solution of the offset constant $D$, its random error in the $\chi^{2}$ fitting
is considerably small ($\pm 1$~m~s$^{-1}$), reflecting the large number of sample points 
($N = 3456$). It should be noted, however, that much larger systematic errors are already 
involved in the absolute values of the original $v_{\rm obs}$ data, especially due to the ambiguity 
in the adopted formula of the $\mu$-dependent convective blue shift given by Equation~3,
which are estimated to be on the order of several tens m~s$^{-1}$ to $\lesssim 100$~m~s$^{1}$
(see Section~6 in Paper~II). Accordingly, the $D$ value of 675~m~s$^{-1}$ derived from  
this analysis (similar to the value of 698~m~s$^{-1}$ obtained in Paper~II)
is regarded as reasonable in comparison with the true value of the gravitational
redshift (633~m~s$^{-1}$).

\subsection{Case of Dividing $\psi$ Into Narrow Bins}

Since we may set $v_{\rm obs} = V_{\rm rot}$ (+ const.), the relation
$v_{\rm obs}/\cos B_{0} = \omega(\psi) x \;\; {\rm (+ const.)}$
holds according to Equation~8.
Let us consider the case where sample points are in a narrow range of
$\psi$ and thus their $\omega$'s are similar to each other. 
Then, $v_{\rm obs}/\cos B_{0}$ may be regarded as linearly 
dependent upon $x$ and its slope (proportionality constant) gives
$\omega$ for this group of data. 

Following this idea, all data points are divided according to 
the values of $\psi$ into fifteen $10^{\circ}$-bins (centered at
$-70^{\circ}$, $-60^{\circ}$, $\ldots$, $0^{\circ}$, $\ldots$, 
$+60^{\circ}$, and $+70^{\circ}$).
Expressing $v_{\rm obs}/\cos B_{0}$ by a linear function $p + qx$, 
\begin{equation}
\chi^{2} = \sum_{i=1}^{N}\Biggl[\frac{v_{{\rm obs},i}/\cos B_{0} - (p + q x_{i})}{\sigma_{i}}\Biggr]^{2}  
\end{equation}
was computed for each bin, where $\sigma_{i} \equiv \epsilon_{i}/\cos B_{0}$, 
$p$ is a constant reflecting the gravitational redshift ($D/\cos B_{0}$), 
and $q$ is equivalent to $\omega$.
The parameters $p$ and $q$ were determined by minimizing $\chi^{2}$ as done 
in Section~5.1. The results derived for each of the 15 $\psi$ bins 
are summarized in Table~1. The $v_{\rm obs}$ versus $x$ plots for each 
group are illustrated in Figure~4, where the linear-regression lines
corresponding to the solutions of $p$ and $q$ are also drawn.    

\setcounter{table}{0}
\begin{table}[h]
\small
\caption{Results of angular velocity analysis at each latitude bin.}
\begin{center}
\begin{tabular}
{cccccc}\hline 
$\psi_{\rm m}$ & [$\psi_{\rm l}, \psi_{\rm u}$] & $N$ & $p$ & $q \times R^{\dagger}$ & $\omega$ \\
(deg) & (deg) &   &  (km~s$^{-1}$) & (km~s$^{-1}$) & (deg~day$^{-1}$) \\
\hline
$-70$ & $[-75, -65]$ &   3 & $0.633 (\pm 0.064)$ & $1.589 (\pm 0.655)$ & $11.312 (\pm  4.662)$ \\
$-60$ & $[-65, -55]$ &  30 & $0.700 (\pm 0.015)$ & $1.739 (\pm 0.083)$ & $12.381 (\pm  0.589)$ \\
$-50$ & $[-55, -45]$ &  70 & $0.723 (\pm 0.010)$ & $1.702 (\pm 0.031)$ & $12.113 (\pm  0.222)$ \\
$-40$ & $[-45, -35]$ & 120 & $0.714 (\pm 0.007)$ & $1.824 (\pm 0.018)$ & $12.985 (\pm  0.132)$ \\
$-30$ & $[-35, -25]$ & 174 & $0.708 (\pm 0.006)$ & $1.893 (\pm 0.012)$ & $13.470 (\pm  0.084)$ \\
$-20$ & $[-25, -15]$ & 264 & $0.698 (\pm 0.005)$ & $1.963 (\pm 0.010)$ & $13.970 (\pm  0.068)$ \\
$-10$ & $[-15, -5]$  & 366 & $0.682 (\pm 0.004)$ & $1.952 (\pm 0.008)$ & $13.892 (\pm  0.055)$ \\
$ +0$ & $[-5, +5]$   & 851 & $0.666 (\pm 0.003)$ & $1.939 (\pm 0.006)$ & $13.803 (\pm  0.046)$ \\
$+10$ & $[+5, +15]$  & 607 & $0.681 (\pm 0.003)$ & $1.964 (\pm 0.007)$ & $13.976 (\pm  0.052)$ \\
$+20$ & $[+15, +25]$ & 358 & $0.669 (\pm 0.004)$ & $1.919 (\pm 0.008)$ & $13.656 (\pm  0.056)$ \\
$+30$ & $[+25, +35]$ & 256 & $0.660 (\pm 0.005)$ & $1.860 (\pm 0.010)$ & $13.240 (\pm  0.072)$ \\
$+40$ & $[+35, +45]$ & 170 & $0.671 (\pm 0.006)$ & $1.753 (\pm 0.013)$ & $12.479 (\pm  0.095)$ \\
$+50$ & $[+45, +55]$ & 102 & $0.662 (\pm 0.008)$ & $1.714 (\pm 0.023)$ & $12.200 (\pm  0.165)$ \\
$+60$ & $[+55, +65]$ &  54 & $0.669 (\pm 0.011)$ & $1.600 (\pm 0.037)$ & $11.390 (\pm  0.261)$ \\
$+70$ & $[+65, +75]$ &  28 & $0.691 (\pm 0.014)$ & $1.707 (\pm 0.085)$ & $12.152 (\pm  0.607)$ \\
\hline
\end{tabular}
\end{center}
Note.\\
$\psi_{\rm m}$, $\psi_{\rm l}$, and $\psi_{\rm u}$ are the middle value, lower boundary, and
upper boundary of each $\psi$ bin. $N$ is the number of data included in each bin.
$p$ and $q$ are the results of least-square analysis, and $\omega$ (deg~day$^{-1}$)
 is derived as $q$ (s$^{-1}$) $\times (180/\pi)\times 24 \times 60 \times60$. \\
$^{\dagger}$$R (= 6.955 \times 10^{5})$ is the solar radius in km. 
\end{table}

\setcounter{figure}{3}
\begin{figure}
\centerline{\includegraphics[width=8.0cm]{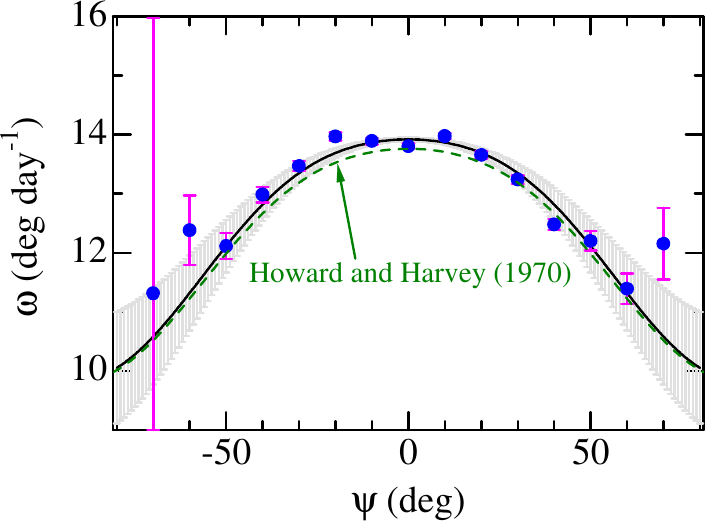}}
\caption{
The finally obtained $\omega$ versus $\psi$ relation, 
$\omega$ = $13.92 (\pm 0.03)$ $-1.69(\pm 0.34)\sin^{2}\psi$ 
$-2.37(\pm 0.62) \sin^{4}\psi$, is shown by the solid line,
where the uncertainty range (corresponding to the errors
in $A$, $B$, and $C$) is indicated by the gray band.
The alternative $\omega$ solutions derived for 15 $\psi$ bins
of 10$^{\circ}$ wide (cf. Table~1) are also overplotted by 
blue filled circles (along with error bars).  
In addition, the $\omega(\psi)$ curve based on Mt. Wilson 
observations by Howard and Harvey (1970) is depicted by the 
dashed line for comparison.
}
\label{fig4}
\end{figure}

\setcounter{figure}{4}
\begin{figure}
\centerline{\includegraphics[width=9.0cm]{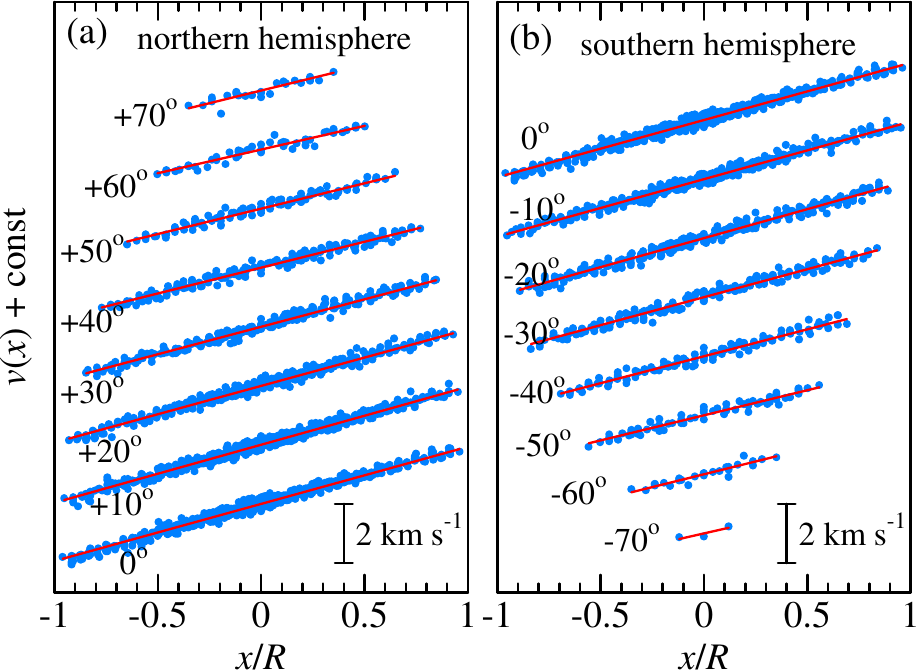}}
\caption{The observed $v_{\rm obs}$ values belonging to each of the 
15 $\psi$-bins (cf. Table~1) are plotted against $x$,
where the corresponding linear regression lines (see Table~1 for 
the best-fit values of $p$ and $q$ determined by the least-squares 
analysis) are also overplotted.
The results for each $\psi$-bin are vertically shifted by   
2~km~s$^{-1}$ relative to the adjacent ones. 
(a) Northern hemisphere ($\psi$ from $0^{\circ}$ to $+70^{\circ}$).
(b) Southern hemisphere ($\psi$ from $0^{\circ}$ to $-70^{\circ}$).
}
\label{fig5}
\end{figure}

The resulting $\omega$ values are plotted against $\psi$ in Figure~5 (symbols),
in order to compare with the $\omega(\psi)$ curve determined in Section~5.1 
(solid line). We can recognize from this figure that both are in satisfactory 
agreement at low-to-middle latitude ($-50^{\circ} \lesssim \psi \lesssim +50^{\circ}$). 
However, appreciable discrepancies are seen at higher latitude of 
$|\psi| \approx$~60\,--\,70$^{\circ}$ where the number of available points is small,
which implies that using sufficiently numerous data points is an 
important factor for precisely determining $\omega$ by this method.

\section{Testing Various Distributions of Observing Points}

Now that we have obtained the rotational parameters based on all 3456 points
covering the solar disk (which are regarded as the ``standard solutions''), 
it may be meaningful to examine how the results are affected in cases of 
different distributions (where data points are reduced in various ways),
by which information of the critical factor(s) may be obtained.

For this purpose, we try three types of test distributions as illustrated 
in Figure~6, and the results to be compared with the standard ``all data'' 
case are presented in Table~2 and Figure~7.  Each of the experiments are 
described below.

\setcounter{figure}{5}
\begin{figure}
\centerline{\includegraphics[width=8.0cm]{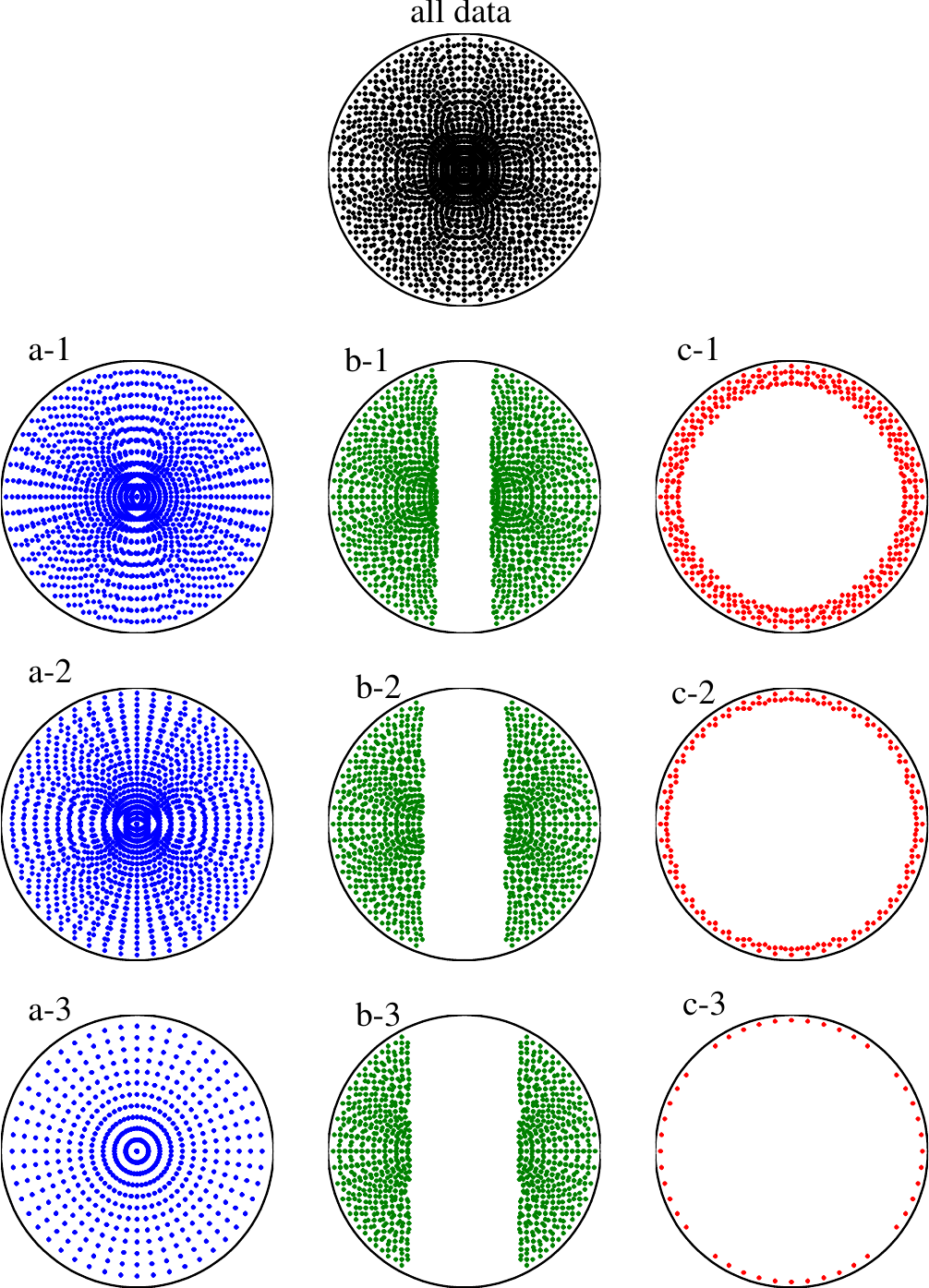}}
\caption{
Illustrated are the distributions of the data points on the disk, 
which were adopted for each of the nine test cases described in Table~2.
The standard ``all data'' case is also shown at the top for comparison.
}
\label{fig6}
\end{figure}

\setcounter{figure}{6}
\begin{figure}
\centerline{\includegraphics[width=8.0cm]{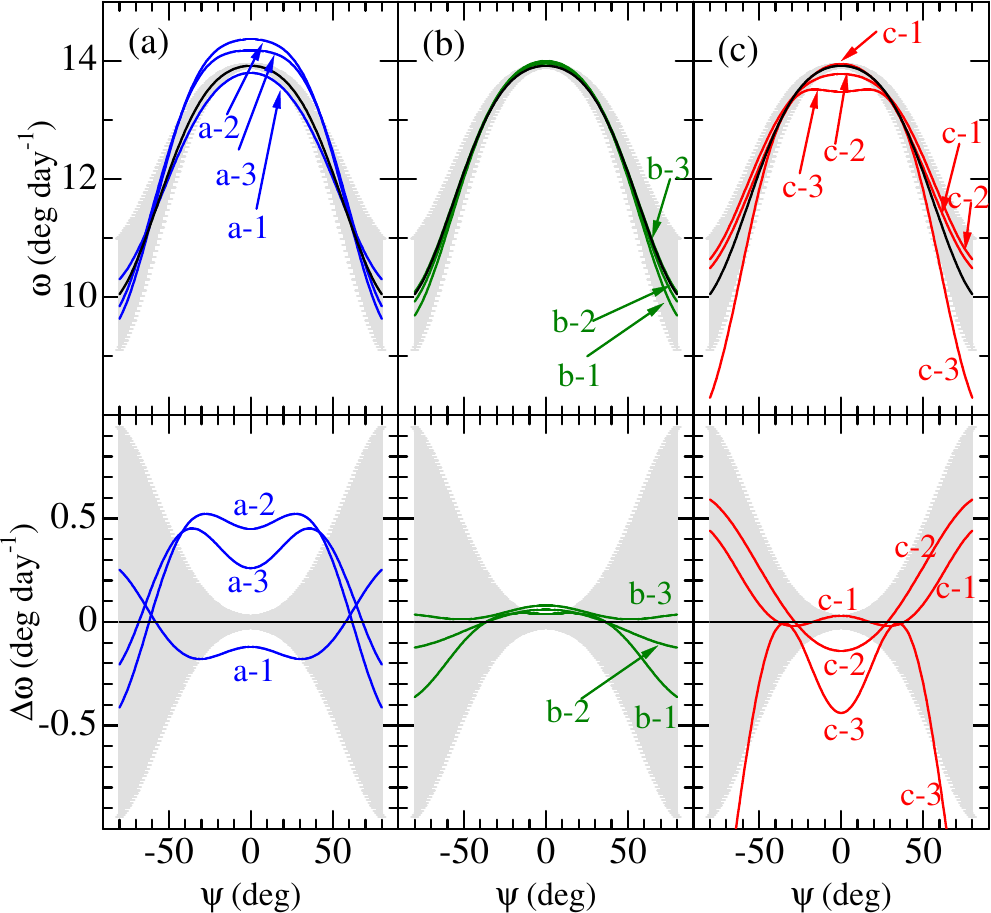}}
\caption{Graphical display for the results of nine test calculations 
(with variously reduced points on the disk, cf. Table~2).
(a) Cases of a-1, a-2, and a-3 (in blue).
(b) Cases of b-1, b-2, and b-3 (in green). (c) Cases of c-1, c-2, 
and c-3 (in red). The result for the standard ``all data'' case is also 
shown by the black line 
(with the uncertainty region depicted as the gray band as in Figure~4) for comparison. 
In each of (a), (b), and (c), the upper panel shows the $\omega$ versus $\psi$ 
relations, while the differences from the ``all data'' case ($\Delta \omega$) are 
plotted against $\psi$ in the lower panel.    
}
\label{fig7}
\end{figure}

\subsection{Case (a): Full Disk But with Reduced Density}

This test is to examine the impact of reducing the number of points 
while retaining the fill-disk covering feature. For this purpose,
out of the 6 types of sub-spectra (Al, Ac, Ar, Bt, Bm, Bb; see Section~2), 
(1) only A-set (Al, Ac, Ar), (2) only B-set (Bt, Bm, Bb), and
(3) only Ac and Bm, were selected, each of which are called
a-1, a-2, and a-3, respectively.

It is interesting to observe from Figure~7a that, while the result for a-1 
is reasonably consistent (in the overall sense) with that of all data, those 
for a-2 and a-3 show appreciable differences despite that their distributions 
also cover the entire disk (cf. Figure~6). Therefore, although higher density of
observing points should generally be more preferable, it is necessary that 
the points are densely distributed in the E--W direction (like a-1) 
but not necessarily in the N--S direction (like a-2).  
 
\subsection{Case (b): Rejecting Points of Near-Meridian Area}

This is the test example where any points whose $|x|$ values (distance from 
the meridian) within a certain threshold are excluded, which corresponds to 
the treatment done in Paper~I in deriving the final $\omega$ versus $\psi$ relation 
(because of the severely enhanced $\omega$ errors at small $|x|$, cf. Section~1.2.2). 
Three cases of $|x|/R> 0.2$ (b-1),  $|x|/R> 0.3$ (b-2, same as in Paper~I), 
and $|x|/R> 0.4$ (b-3) were tried.

As shown in Figure~7b, the differences from the standard ``all data'' result 
turned out insignificant in all of these three cases (b-1, b-2, b-3). 
This presumably suggests that the behaviors of $\omega$ are primarily 
determined by the off-meridian region of larger $|x|$ where the contribution 
of $\omega$ to $v_{\rm obs} (\approx \omega x)$ is more significant.
Therefore, we may state that the choice of excluding $|x| < 0.3R$ in Paper~I 
was reasonable (though not well recognized at that time), which may explain 
the consistency between the results in Paper~I and this study (cf. Section~5.1).

\subsection{Case (c): Only Circular Region Along the Limb}

This test is to simulate the case where the observed points are distributed 
only in the narrow circular band around the limb. As a matter of fact, 
quite a few spectroscopic determinations of solar rotation (in particular, 
old ones) relied upon the spectra only near to the limb because the Doppler 
shift is the largest and easier to detect there.
Here, three cases of $r/R> 0.8$ (c-1),  $r/R> 0.9$ (c-2), 
and $r/R> 0.95$ (c-3) are examined. 

Figure~7c elucidates that the $\omega(\psi)$ results for these cases 
do not match well that of the standard case (especially at higher $\psi$). 
Although c-1 may still be regarded as in favorable agreement at lower 
$\psi$ (though not good at higher $\psi$), the situation becomes 
worse at c-2 and c-3 (especially c-3, the most outer distribution, shows 
a considerable discrepancy).
This is presumably related to the fact these Case (c) tests (in particular 
c-3) are based on much smaller number of points ($N$) compared to Cases (a) 
and (b) (see Table~2).

\setcounter{table}{1}
\overfullrule=0pt
\begin{table}[h]
\scriptsize
\caption{Tests of how the results are affected by variously reducing the observed points on the disk.}
\begin{center}
\begin{tabular}
{ccccccl}\hline 
   & $N$ & $A$ & $B$ & $C$ & $D$ & Remark \\
\hline
all data & 3456 & $13.92 (\pm 0.03)$ & $ -1.69 (\pm 0.34)$ & $ -2.37 (\pm 0.62)$ & $ 0.675 (\pm 0.001)$ &  (Al, Ac, Ar, Bt, Bm, Bb) \\
\hline
\multicolumn{7}{c}{(Data of specific slit positions are excluded)}\\
a-1  & 1728 & $13.80 (\pm 0.04)$ & $ -2.14 (\pm 0.39)$ & $ -1.51 (\pm 0.72)$ & $ 0.671 (\pm 0.002)$ &  only (Al, Ac, Ar) \\
a-2  & 1728 & $14.37 (\pm 0.06)$ & $ -1.00 (\pm 0.66)$ & $ -4.00 (\pm 1.18)$ & $ 0.692 (\pm 0.002)$ &  only (Bt, Bm, Bb) \\
a-3  & 1152 & $14.18 (\pm 0.06)$ & $ -0.56 (\pm 0.68)$ & $ -4.03 (\pm 1.21)$ & $ 0.707 (\pm 0.002)$ &  only (Ac, Bm) \\
\hline
\multicolumn{7}{c}{(Data within a given distance from the meridian $(|x|)$ are excluded)}\\
b-1  & 1886 & $13.96 (\pm 0.03)$ & $ -1.63 (\pm 0.35)$ & $ -2.86 (\pm 0.65)$ & $ 0.669 (\pm 0.002)$ & only $|x|/R > 0.2$\\
b-2  & 1430 & $13.98 (\pm 0.03)$ & $ -1.85 (\pm 0.37)$ & $ -2.40 (\pm 0.70)$ & $ 0.666 (\pm 0.002)$ & only $|x|/R > 0.3$\\
b-3  & 1082 & $14.00 (\pm 0.03)$ & $ -1.91 (\pm 0.41)$ & $ -2.19 (\pm 0.85)$ & $ 0.662 (\pm 0.002)$ & only $|x|/R > 0.4$\\
\hline
\multicolumn{7}{c}{(Data within a given distance from the disk center $(r)$ are excluded)}\\
c-1  &  532 & $13.95 (\pm 0.05)$ & $ -2.10 (\pm 0.44)$ & $ -1.51 (\pm 0.74)$ & $ 0.665 (\pm 0.003)$ & only $r/R > 0.8$\\
c-2  &  220 & $13.78 (\pm 0.07)$ & $ -1.10 (\pm 0.59)$ & $ -2.20 (\pm 0.93)$ & $ 0.668 (\pm 0.005)$ & only $r/R > 0.9$\\
c-3  &   44 & $13.48 (\pm 0.11)$ & $  1.00 (\pm 0.87)$ & $ -6.54 (\pm 1.69)$ & $ 0.649 (\pm 0.012)$ & only $r/R > 0.95$\\
\hline
\end{tabular}
\end{center}
Note.\\
$N$ is the total number of adopted data points (see also footnote~2).
$A$, $B$, and $C$ are in unit of deg~day$^{-1}$, while $D$ is in km~s$^{-1}$. See Figure~6 
for the graphical display of the distributions for each of these ten cases.   
\end{table}

\section{Difficulty of Spectroscopic Solar Rotation Measurement}

The test simulations done in Section~6 have shown that 
spectroscopically determined solar rotation parameters (not only
the $\psi$-dependent nature of differential rotation but also the
equatorial rotation velocity at $\psi = 0^{\circ}$) depend rather 
significantly upon how the observed points on the disk are chosen.

Presumably, this is attributed to the irregularly fluctuating velocity 
components in the solar photosphere. If the velocity field on the solar disk
were steady and monotonic, reliable measurement of the rotation law would be
feasible based on not so many observing points. However, the actual
surface of the Sun is violently in motion and covered with inhomogeneous 
velocity fields as demonstrated in Figure~8,  
where the 3D representations of (a) $v_{\rm obs}(x,y)$ and 
(b) $\Delta v(x,y)$ (residual of $v_{\rm obs}$ after subtracting the 
rotational velocity field $V_{\rm rot}$) are depicted based on our data.
As seen from Figure~8b, $\Delta v$ shows a random variation 
on the order of $\approx 10^{2}$~m~s$^{-1}$. Therefore, if the distribution 
of observing points is not so dense as to sufficiently cancel out this 
inhomogeneous velocity fields, incorrect rotation parameters may result.      

\setcounter{figure}{7}
\begin{figure}
\centerline{\includegraphics[width=12.0cm]{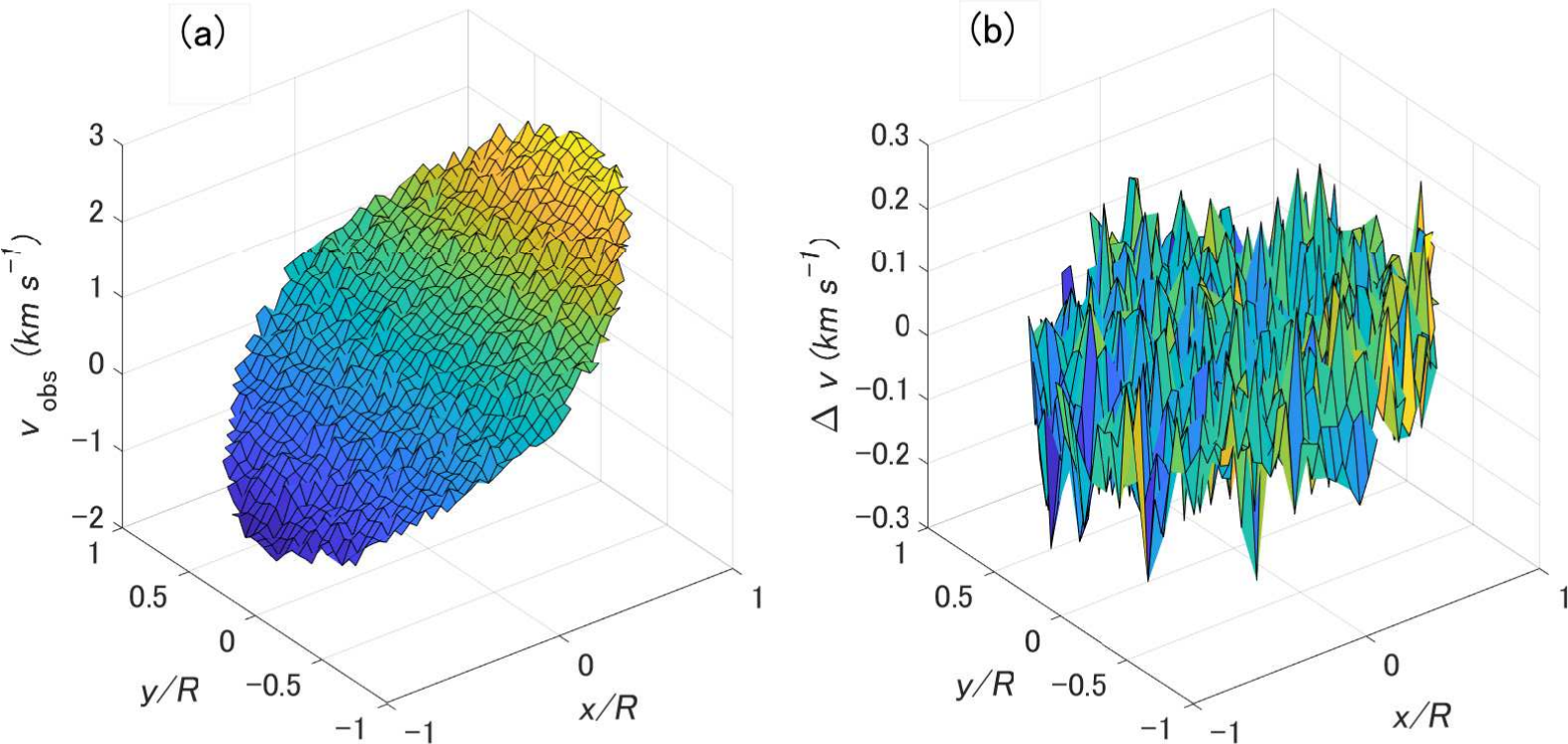}}
\caption{3D representation of (a) $v_{\rm obs}(x,y)$
and (b) $\Delta v(x,y)$, where the $x$--$y$ plane is divided into 
$40 \times 40$ square segments. Here, $\Delta v$ is the 
difference between $v_{\rm obs}$ and $V_{\rm rot}$ 
(rotational velocity field corresponding to the final solutions 
of $A$, $B$, $C$, and $D$).
}
\label{fig8}
\end{figure}

Accordingly, it is likely that some previously published solar rotation results 
obtained by Doppler shift measurements (especially those based on comparatively 
smaller number of observing points) suffered appreciable uncertainties. 
This may be the cause for the large spread in $\omega$ versus $\psi$ 
curves or some apparent outlier values of A, B, C seen in the past historical 
publications (see, e.g., Figure~2 in Patern\`{o}, 2010, or Figure~12 in Paper~I). 
For the same reason, the suspected cyclic variation of solar equatorial rotation 
velocity (by $\approx \pm 5\%$) with a period of 34 years, which was reported
by Belvedere and Patern\`{o} (1975) based on the historical publications 
from 1900 to 1970, had better be viewed with caution. 

Nevertheless, the spectroscopic rotational parameters in the ``approximate'' 
sense are regarded as being almost established at $A\approx 14$, 
$B \approx -2$, and $C \approx -2$ based on the literature results 
(see Figure~12 in Paper~I). Yet, what is required for proceeding to
the next step (e.g., to detect the dependence of rotation upon the activity 
phase as mentioned in Section~1) should be to further improve the precision 
of determination. Though this is not an easy task
for which nobody knows any definite recipe, what should be generally kept 
in mind may be summarized into the following two points.
\begin{itemize}
\item
First, it is essentially important to employ spectra obtained at as many
(i.e., densely distributed) observing points as possible. Regarding their 
distribution, more weight (higher density) should be given to larger 
$x$ region as well as to E--W direction according to the results in Section~6.
\item
Second, even if using many points on the disk could be accomplished,
only one set of data acquired in a short time
(e.g., in $\approx 1$~day) is not sufficient to warrant the reliability
of the solution. Comparing the results derived by analyzing several or 
more sets of data based on repeated observations over a certain period 
would help to estimate the real precision quantitatively.
\end{itemize}

\section{Summary and Conclusion}

In studying the solar differential rotation on the visible surface, 
two representative methods have been traditionally employed: (i) 
tracing of active regions (such as sunspots) over the rotation cycle 
and (ii) spectroscopic Doppler shift measurement on the disk. 

Following the latter spectroscopic approach, Takeda and Ueno once
investigated in Paper~I the nature of solar differential rotation 
based on the full-disk observations by applying the I$_{2}$-cell technique, 
which is known to be efficient in precise determinations of Doppler 
velocity shifts.

As viewed from the present knowledge, however, the procedure of analyzing 
I$_{2}$+solar composite spectrum adopted in that paper was problematic 
in the sense that the center--limb variation of spectral line strengths 
was not taken into consideration, because the disk-center spectrum was 
exclusively used as the reference spectrum. In addition, the derivation of 
$\omega$ done in Paper~I was not necessarily adequate because it was 
directly converted from $v_{\rm obs}$ at each point of the disk, 
which eventually resulted in considerable errors in $\omega$ at smaller 
$x$ near to the meridian. 

The remedy for the former problem is to use an adequately adjusted 
theoretical spectrum as the comparison reference at each point
as done in Paper~II (where the gravitational redshift was determined 
from the meridian data), by which absolute heliocentric radial 
velocity can be obtained.
Meanwhile, the solution for the latter problem is to derive $\omega$
(not one-by-one from each $v_{\rm obs}$ but) from an ``ensemble'' of 
$v_{\rm obs}$ data by applying the least squares analysis. 

Motivated by these considerations, the author decided to redetermine 
the parameters of solar differential rotation based on absolute radial 
velocities, which were derived from the same observational data as used 
in Paper~I but by applying the analysis procedure devised in Paper~II. 

In deriving the $\omega$ versus $\psi$ relation, two different approaches 
were tried. (a) Expressing $\omega$ by 2nd-order polynomial in terms 
of $\sin^{2}\psi$, a least-squares analysis was applied to the
 whole $v_{\rm obs}$ data over the disk to derive $A$, $B$, and $C$ 
(coefficients of the polynomial) and $D$ (gravitational redshift).
(b) All data points were divided according to $\psi$ into fifteen 
$10^{\circ}$-bins, and a linear regression analysis was applied
to $v_{\rm obs}$ data belonging to each $\psi$-bin to determine
the corresponding $\omega$. 

The first approach resulted in $\omega$ (deg~day$^{-1}$) = $13.92 (\pm 0.03)$ 
$-1.69(\pm 0.34)\sin^{2}\psi$ $-2.37(\pm 0.62) \sin^{4}\psi$ 
along with the gravitational redshift of 675~m~s$^{-1}$. 
Meanwhile, the $\omega$ values derived by the second approach  
at each of the $\psi$-bins are in good agreement at low-to-middle latitude 
($-50^{\circ} \lesssim \psi \lesssim +50^{\circ}$), though appreciable 
discrepancies are seen at higher latitude ($|\psi| \approx$~60\,--\,70$^{\circ}$) 
where the number of available points is small.

This $\omega$ versus $\psi$ relation is almost consistent with that obtained 
in Paper~I, which means that the changes in the results are insignificant 
despite of the updated procedures in this new analysis. 
Likewise, these $A$, $B$, $C$ values are favorably compared with those of 
previous publications (see Figure~12 in Paper~I). 

As a related application of this analysis, the impact of reducing the number 
of observing points was also examined. This test revealed that significant 
changes of rotation parameters are observed in some cases, which is presumably 
due to the irregular velocity field on the solar surface. 
It is necessary, therefore, to pay attention to secure a sufficiently large
number of points on the disk, in order to obtain reliable results of
higher precision. 

\begin{acks}
This investigation has made use of the data obtained by using the 
Domeless Solar Telescope at Hida Observatory of Kyoto University.
\end{acks}



\begin{dataavailability}
The radial velocity data and related quantities at each point of the solar disk, 
upon which this study is based, are available as supplementary material 
online (ReadMe.txt, tableE.dat).
\end{dataavailability}

%


\begin{ethics}
\begin{conflict}
The author declares that he has no conflicts of interest.
\end{conflict}
\end{ethics}


\begin{thebibliography}{spr-mp-sola.bst}
\bibitem[Belvedere&Paterno(1975)]{Belvedere1975}
  Belvedere,~G., Patern\`{o},~L.:1975, \solphys{} \textbf{41}, 289.
  DOI: 10.1007/BF00154066 
\bibitem[Bevington&Robinson(2003)]{Bevington2003}
  Bevington,~P.R, Robinson,~D.K.: 2003, 
  Data Reduction and Error Analysis for the Physical Sciences, 3rd ed., 
  McGraw Hill, New York.
\bibitem[Howard&Harvey(1970)]{Howard1970}
  Howard,~R., Harvey,~J.: 1970, \solphys{} \textbf{12}, 23.
  DOI: 10.1007/BF02276562 
\bibitem[Paterno(2010)]{Paterno2010}
  Patern\`{o},~L.: 2010, \apss{} \textbf{328}, 269.
  DOI: 10.1007/s10509-009-0218-0 
\bibitem[Schroter(1985)]{Schroter1985}
  Schr\"{o}ter,~E.H.: 1985, \solphys{} \textbf{100}, 141.
  DOI: 10.1007/BF00158426 
\bibitem[Takeda(2022)]{Takeda2022}
  Takeda,~Y.: 2022, \solphys{} \textbf{297}, 4. 
  DOI: 10.1007/s11207-021-01931-0 
\bibitem[Takeda&Ueno(2011)]{Takeda2011}
  Takeda,~Y., Ueno,~S.: 2011, \solphys{} \textbf{270}, 447 (Paper~I). 
  DOI: 10.1007/s11207-011-9765-y 
\bibitem[Takeda&Ueno(2012)]{Takeda2012}
  Takeda,~Y., Ueno,~S.: 2012, \solphys{} \textbf{281}, 551 (Paper~II). 
  DOI: 10.1007/s11207-012-0068-8 
\bibitem[Takeda&UeNo(2019)]{Takeda2019}
  Takeda, Y., UeNo, S.: 2019, \solphys{} \textbf{294}, 63.
  DOI: 10.1007/s11207-019-1455-1 
\end{thebibliography}
\end{document}